\documentclass[11pt]{article}
\usepackage[margin=1in]{geometry}
\usepackage{graphicx}

\def\beq{\begin{equation}}
\def\eeq#1{\label{#1}\end{equation}}
\def\eeqn{\end{equation}}

\def\beqa{\begin{eqnarray}}
\def\eeqa#1{\label{#1}\end{eqnarray}}
\def\eeqan{\end{eqnarray}}

\let\bar=\overbar

\def\Dslash{\not{\hbox{\kern-4pt $D$}}}
\def\dslash{\not{\hbox{\kern-2pt $\del$}}}

\def\msb{{\bar{\ssstyle M \kern -1pt S}}}

\def\Title#1{\begin{center} {\Large {\bf #1} } \end{center}}
\def\Author#1{\begin{center} {\normalsize {\sc #1} } \end{center}}
\def\Institution#1{\begin{center} {\normalsize {\it #1} } \end{center}}
\def\Abstract#1{\noindent {\normalsize {\bf Abstract:} {\normalfont #1}}}
\def\Conference{\vspace{4mm}\begin{raggedright} {\normalsize {\it Talk presented at the 2019 Meeting of the Division of Particles and Fields of the American Physical Society (DPF2019), July 29--August 2, 2019, Northeastern University, Boston, C1907293.} } \end{raggedright}\vspace{4mm}}

\begin{document}

%
%

\Title{Searching for Dark Photons with PADME}

\Author{Andre Frankenthal}

\Institution{Department of Physics\\ Cornell University, Ithaca, NY 14853, USA}

\Abstract{PADME is a fixed-target, missing-mass experiment at the Laboratori Nazionali di Frascati (LNF) to search for evidence of dark photons. It has collected a first set of commissioning data in 2018/2019 which will also be used for preliminary data analysis. PADME is expected to reach a sensitivity of up to $10^{-8}$ in $\epsilon^2$ (kinetic mixing coefficient) for low-mass dark photons ($\sim$ few MeV). Here we describe PADME's design, status of the experiment, and future plans.}

\Conference

%
%

\section{Introduction}

The dark photon is a popular hypothetical mechanism to bridge the Standard Model (SM) and the dark sector. It is the mediator of a new $U(1)_D$ gauge symmetry, similar to SM's $U(1)_Y$ \cite{Holdom:1985ag}. This new symmetry provides a renormalizable extension to the SM Lagrangian, where a mixing between the kinetic terms of the $U(1)_D$ and $U(1)_Y$ gauge bosons effectively couples the two sectors \cite{battaglieri2017cosmic}. This coupling is controlled by the kinetic mixing coefficient $\epsilon$, and leads to an electromagnetic current connected to the dark photon particle. Processes such as positron-electron annihilation ($e^+ e^- \rightarrow \gamma A'$), where the second photon converts into a dark photon, are then allowed and predicted by the model. This is the main signature of interest to PADME.

\section{PADME}

\begin{figure}[htb]
\centering
\includegraphics[height=2.5in]{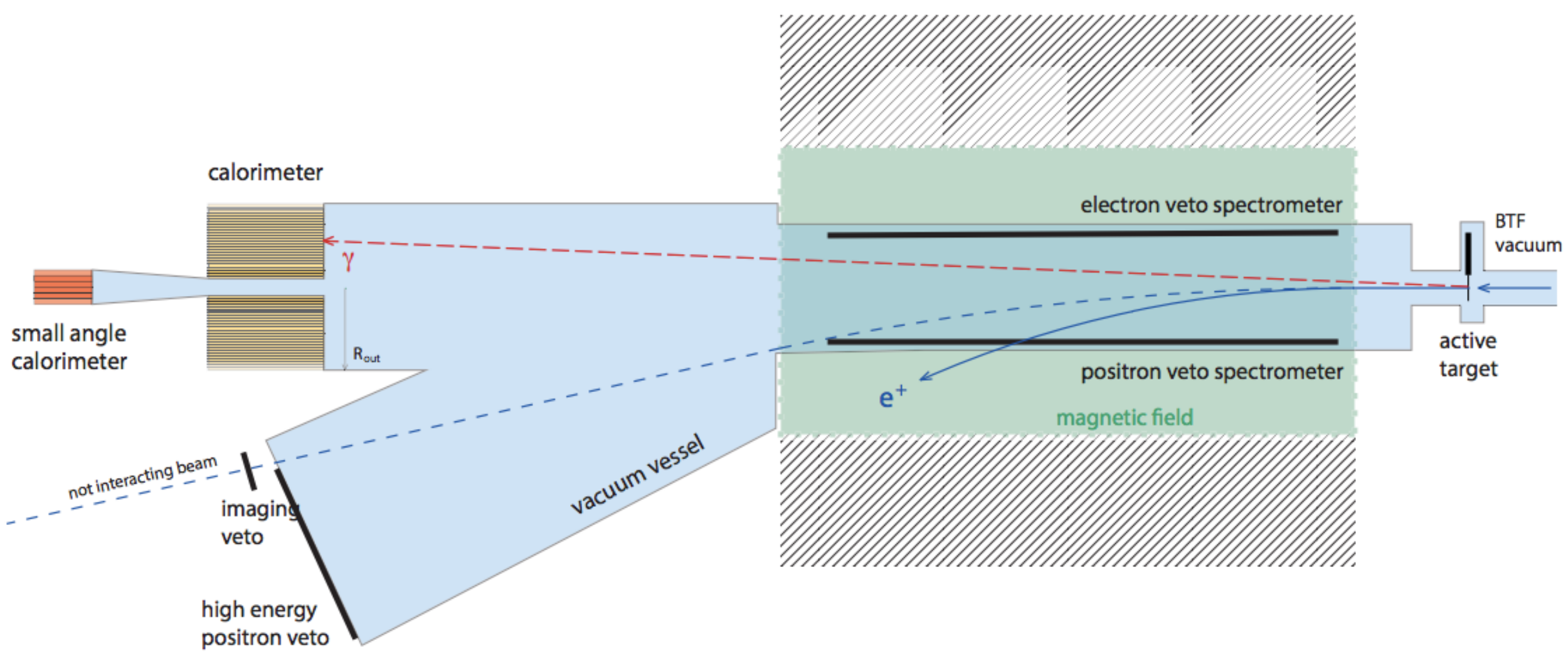}
\caption{Schematic layout of the PADME experiment.}
\label{fig:exp_setup}
\end{figure}

The Positron Annihilation into Dark Matter Experiment (PADME) seeks to find evidence for dark photons from positron-electron annihilation events \cite{Raggi:2014zpa}. The experimental design is shown in Fig.~\ref{fig:exp_setup}.

A beam of positrons produced by the DA$\Phi$NE accelerator at LNF strikes a fixed target made of polycrystalline diamond. If carbon electrons and protons annihilate producing a photon and a dark photon, the latter will escape undetected, while the former will follow a straight trajectory and strike a large electromagnetic calorimeter (ECAL) further downstream. Non-interacting positrons are diverted off the ECAL's path and onto a beam dump by means of a 0.5 T magnet placed immediately downstream of the target. By measuring precisely the energy and impact position of the photon with the ECAL, its momentum can be determined, since it originates in a fixed target. Combining this information with a precise measurement of the beam momentum and the fact that the electrons are bound to the carbon atoms, the missing mass of the event can be inferred:

\begin{figure}[b!]
\centering
\includegraphics[height=3.0in]{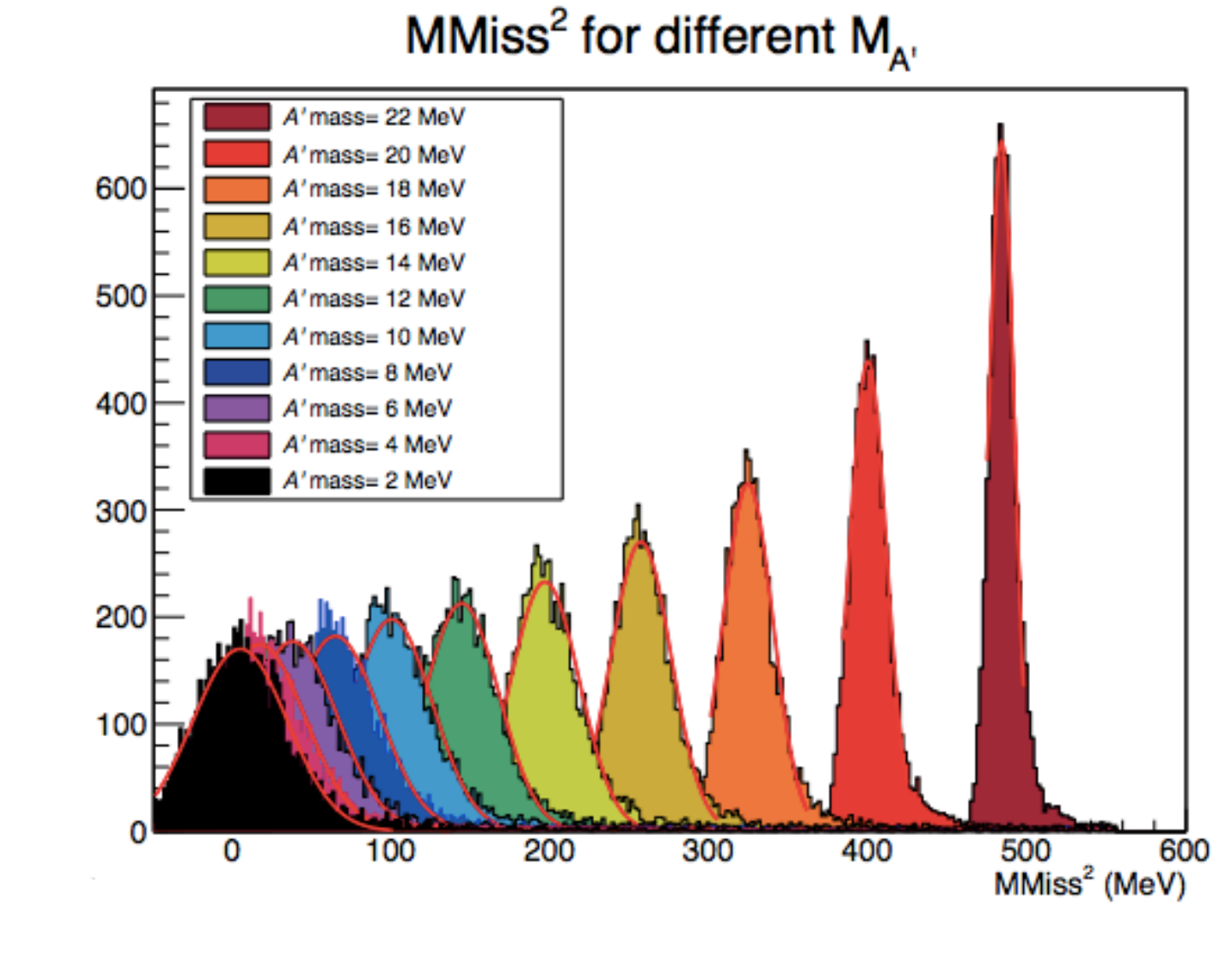}
\caption{Missing mass spectrum expected for different choices of dark photon mass within the detector acceptance, as determined from a Geant4-based Monte Carlo simulation.}
\label{fig:miss_mass}
\end{figure}

\begin{equation}
	m^2_{\textrm{miss}} = (p_{e^-} + p_{e^+} - p_\gamma)^2
\end{equation}

\noindent A Geant4-based Monte-Carlo simulation of the missing mass spectrum expected for different choices of dark photon mass is shown in Fig.~\ref{fig:miss_mass}. 

\begin{figure}[htb]
\centering
\includegraphics[height=2.0in]{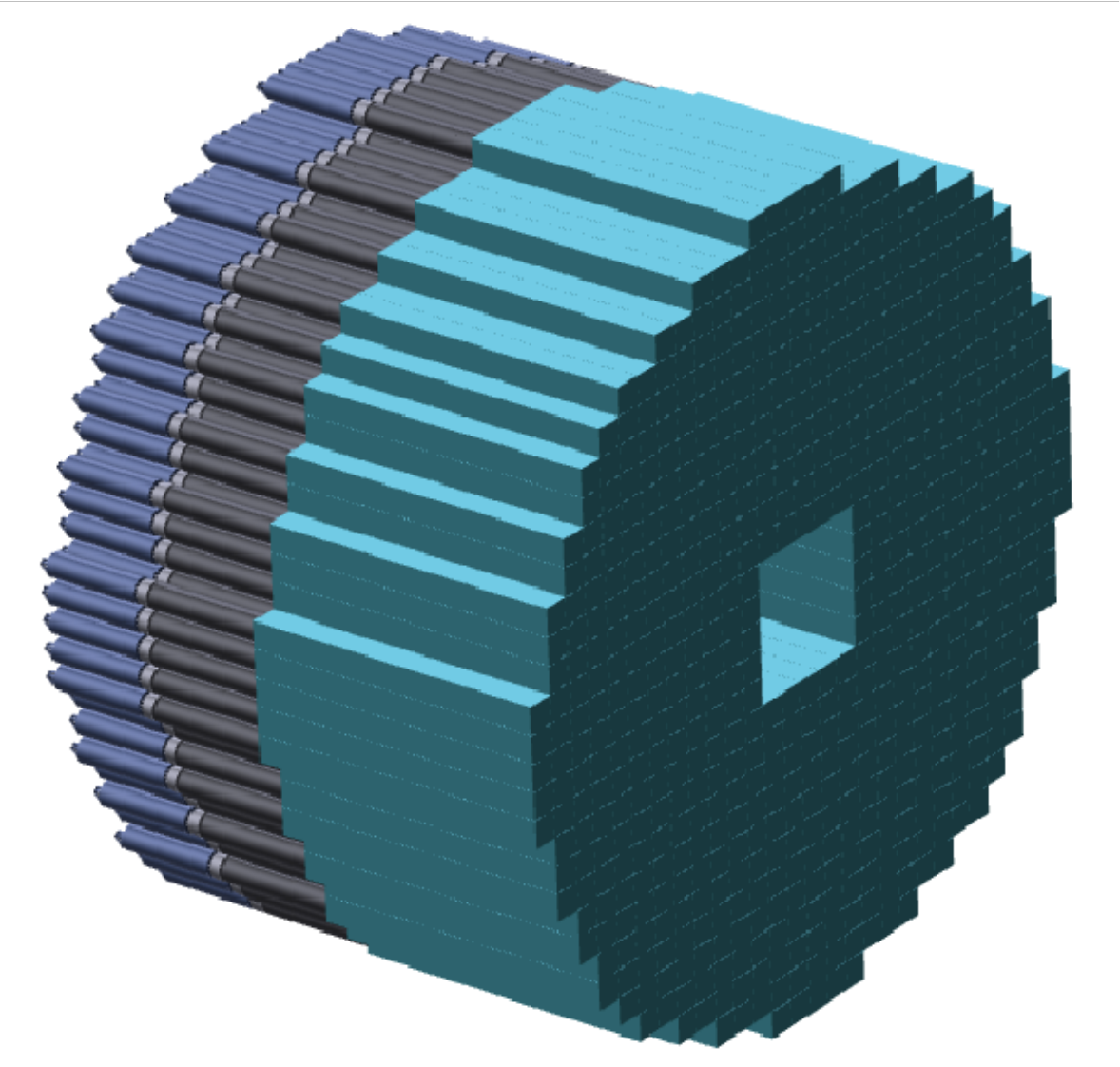}
\caption{Schematic layout of the PADME electromagnetic calorimeter (ECAL), consisting of 616 BGO scintillating crystals with a central square hole for mitigating background Bremsstrahlung photons.}
\label{fig:ecal}
\end{figure}

The ECAL consists of 616 BGO scintillating crystals (from the old L3 experiment at LEP) with dimensions 21x21x230 mm$^3$ each, arranged in a cylindrical orientation. Coupled to each crystal is a HZC photomultiplier tube (PMT), for scintillation light readout. The ECAL is placed roughly 3~m downstream from the target, and thus has an angular acceptance between 20 and 95 mrad. It has an energy resolution of 1-2\%/$\sqrt{E}$ for positrons below a GeV \cite{Raggi:2016ews}. Fig.~\ref{fig:ecal} shows a schematic diagram of the ECAL.

The main backgrounds to the expected signature are classic annihilation ($e^+e^- \rightarrow \gamma\gamma(\gamma)$) and Bremsstrahlung ($e^+N \rightarrow e^+N\gamma$) events. PADME has a range of auxiliary detectors designed to mitigate such backgrounds. The distribution of Bremsstrahlung photons is sharply peaked toward small angles and low energies. To prevent these soft photons from impacting the hard photon energy measurements, the ECAL is built with a central square hole to let the Bremsstrahlung photons pass through. 

Moreover, two additional sub-detectors -- the Positron Veto and Small-Angle Calorimeter (SAC) -- help in the mitigation of Bremsstrahlung events. If a positron radiates a photon, it will lose energy and bend more sharply under PADME's magnetic field. Instead of striking the beam dump, such positrons will hit the walls of the magnet. We therefore instrument the walls of the magnet with plastic scintillators to identify these lower-energy positrons \cite{Oliva:2019pfd}. 

\begin{figure}[htb]
\centering
\includegraphics[height=3.0in]{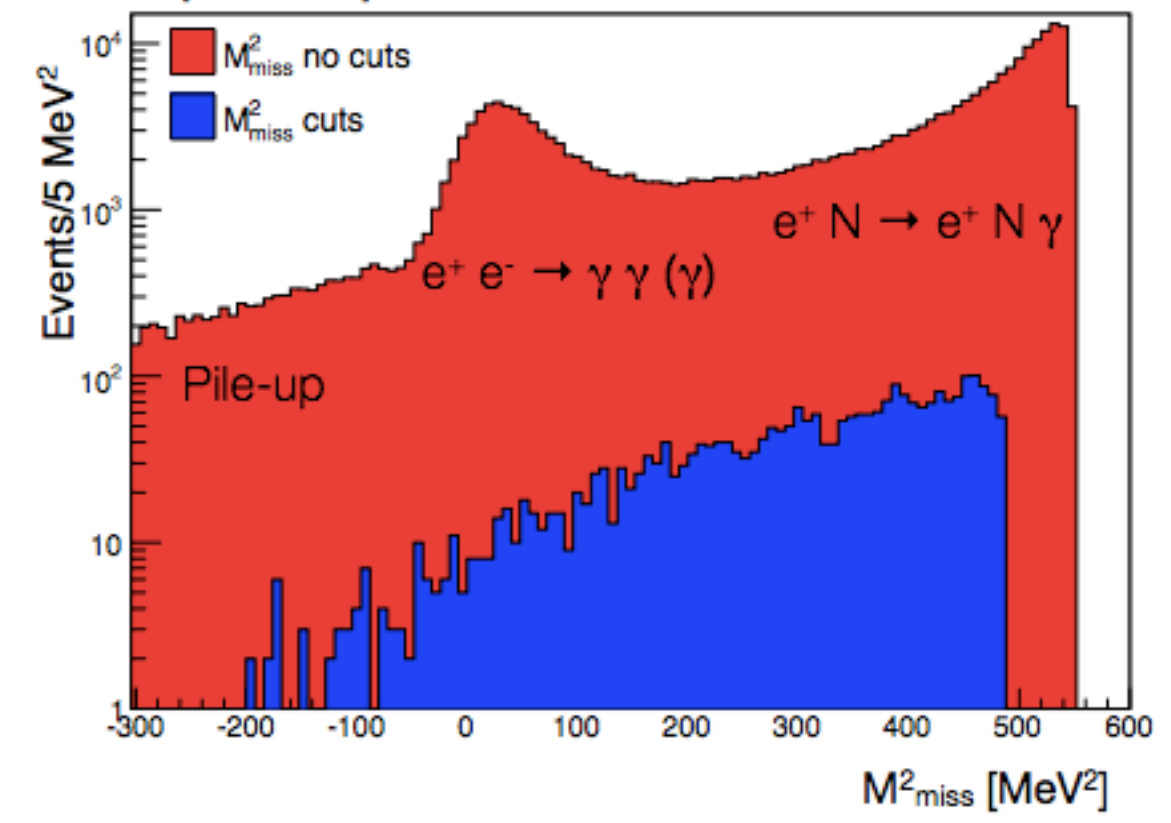}
\caption{Missing mass spectrum PADME backgrounds, before and after the mitigations described in the text, as determined from a Geant4-based Monte Carlo simulation.}
\label{fig:backgrounds}
\end{figure}

The SAC is a small and fast Cherenkov calorimeter, composed of 25 PbF$_2$ crystals and placed downstream of the ECAL, flushed with its central square hole. The fast performance of the SAC allows the individual identification of Bremsstrahlung photons, despite the large rate \cite{Frankenthal:2018yvf}. By performing an in-time correlation between the Positron Veto and the SAC measurements, Bremsstrahlung events can be tagged and rejected. The SAC is also useful in tagging annihilation events. In this case, the in-time correlation between ECAL and SAC photons allows the reconstruction of the original positron beam momentum, and the rejection of such events. Fig.~\ref{fig:backgrounds} shows a Geant4-based MC study of the missing mass spectrum due to backgrounds, before and after the mitigations described above.

\begin{figure}[b!]
\centering
\includegraphics[height=3.0in]{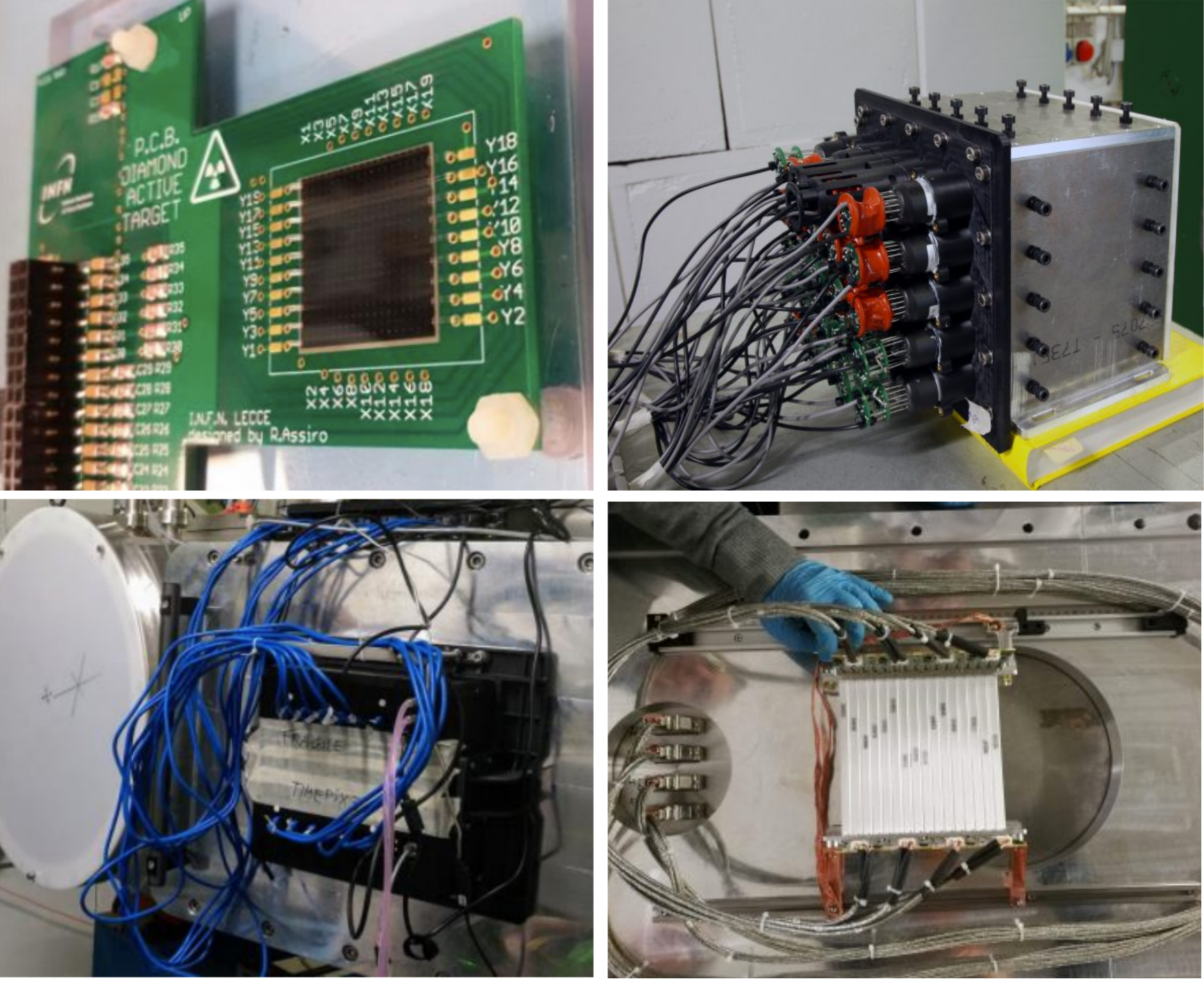}
\caption{Auxiliary detectors used in the PADME experiment. Top left: polycrystalline diamond active target; top right: Small-Angle Calorimeter; bottom right: positron vetoes; bottom left: TimePix3.}
\label{fig:aux_dets}
\end{figure}

Finally, PADME also has two dedicated beam monitors: the TimePix3 detector and the active diamond target. The active target's surface is equipped with 19 graphite strips in each orientation, which collect charge released from the positron beam interactions. The total deposited charge is proportional to the beam intensity, and in addition a coarse beam profile measurement is possible \cite{Oliva:2019alx}. The TimePix3 detector consists of 12 sensors of 256x256 pixels each, with pitch size 55 microns, for close to a million pixels in total. Its purpose is to monitor the non-interacting beam, and measure the positron momentum distribution. Fig.~\ref{fig:aux_dets} shows the four auxiliary detectors.

\section{Preliminary Data and Beam-Induced Backgrounds}

PADME collected its first set of data over 4 months in 2018 and 2019. The data are being used for detector commissioning work as well as preliminary data analysis. From this original set of data, a new source of backgrounds not previously accounted for was identified: beam-induced backgrounds. This leads to a large unexpected deposition of energy in the ECAL. Additional MC studies helped to understand where this extra deposit originates. There are two main sources: the Beryllium window that is used to separate the accelerator vacuum from the detector vacuum, and the magnet that steers the beam from the injection point into the target. 

\begin{figure}[htb]
\centering
\includegraphics[height=2.3in]{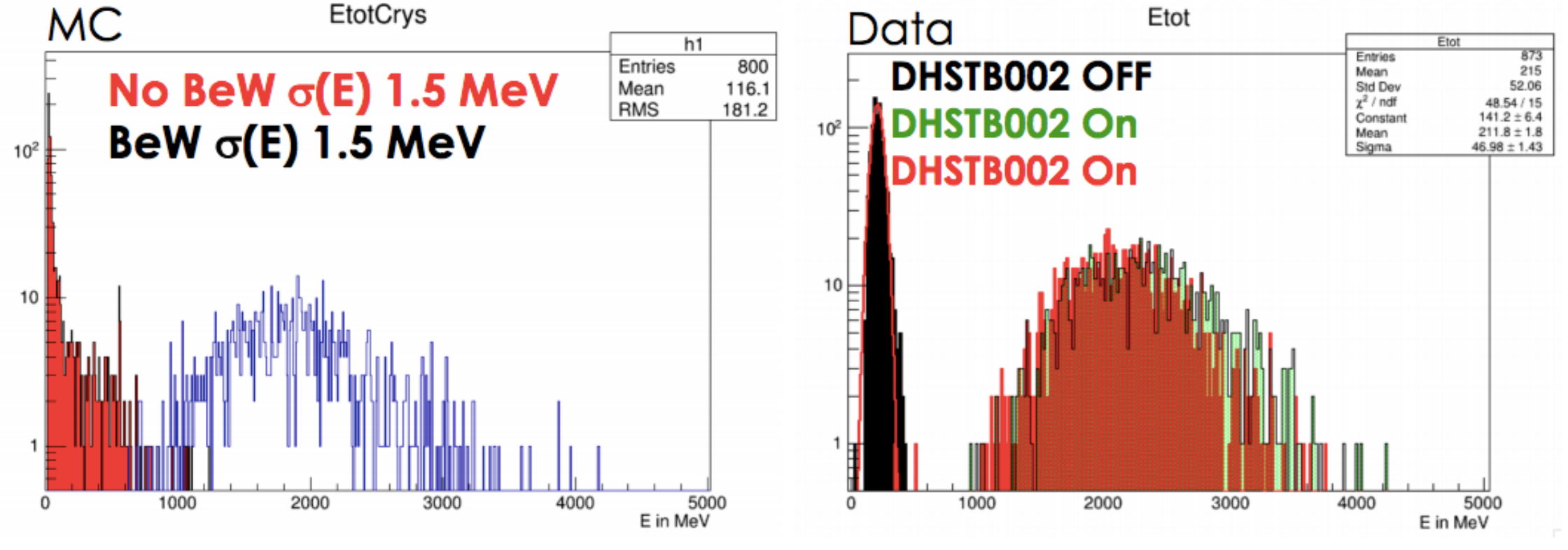}
\caption{Preliminary data and MC energy deposition in the ECAL, due to beam-induced backgrounds. Left: effect of adding to the experiment a Beryllium window (BeW) with beam energy spread of 1.5 MeV, according to MC; right: effect of turning off the steering magnet (DHSTB002) that transports the positron beam onto the target, in data.}
\label{fig:beam_bkgs}
\end{figure}

The Beryllium window causes extra scattering of beam positrons, creating secondary particles that are not well centered with the beam trajectory and therefore eventually strike the ECAL indiscriminately. Furthermore, the beam has a non-zero energy spread, and positrons with an energy too low or too high will leave the steering magnet at an angle, hitting other parts of the detector and showering further secondary particles into the ECAL. Fig.~\ref{fig:beam_bkgs} shows the pattern of energy deposition in the calorimeter, for data (right) and MC (left). The data plot compares configurations with the steering magnet on vs off, while the MC plot contrasts configurations with the Beryllium present vs absent. 

The PADME Collaboration is studying ways to mitigate this background: a few ideas include moving the Beryllium window downstream of the steering magnet, and adding shielding to the front face of the magnet, around the target. These steps should remove a large fraction of beam-induced backgrounds for the next run of PADME.


\section{Status and Future Plans}

\begin{figure}[htb]
\centering
\includegraphics[height=3.5in]{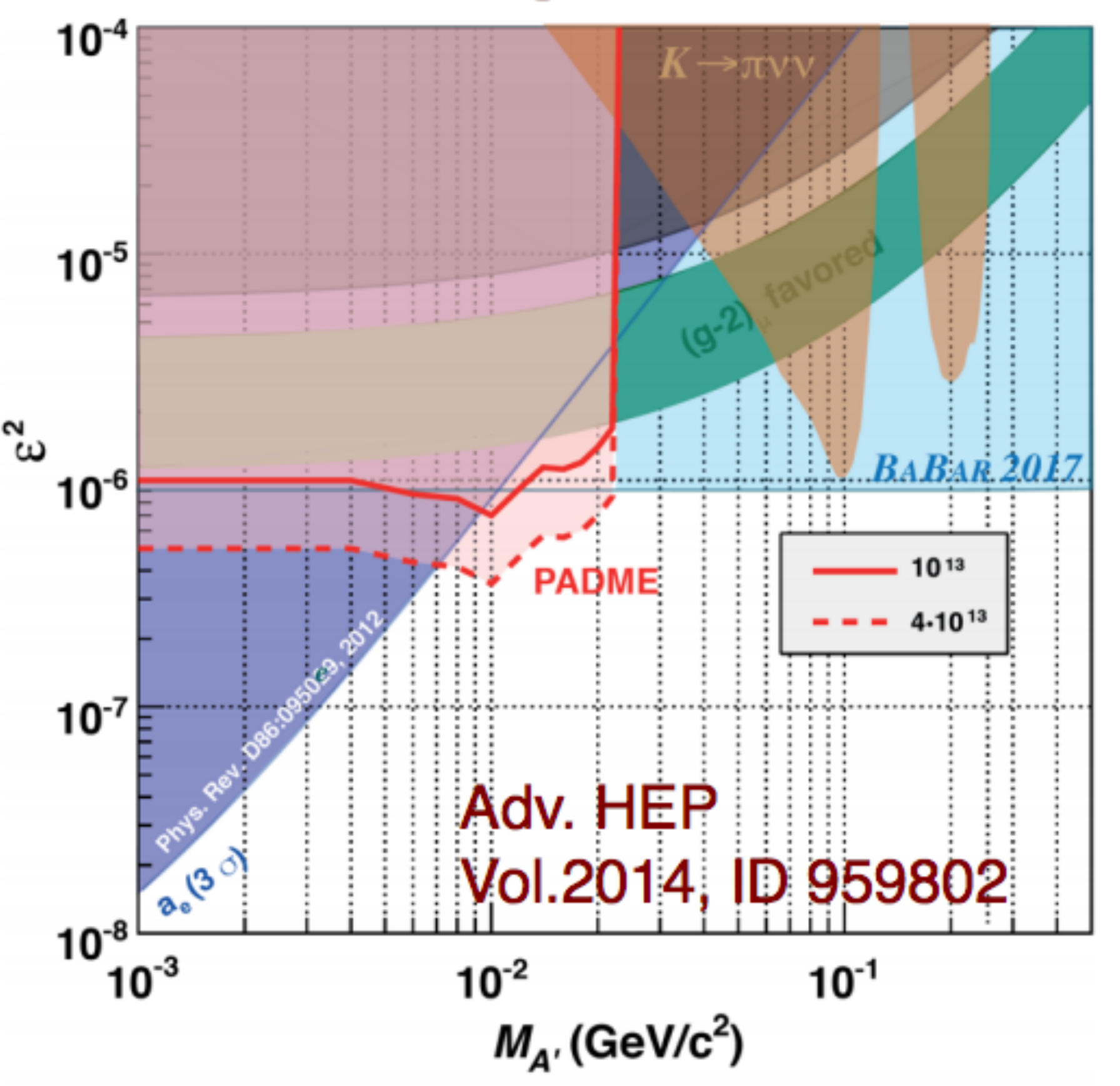}
\caption{Expected PADME sensitivity for 6 months and 2 years of data-taking. The dark photon mass range is constrained by the positron beam energy at LNF, 550 MeV.}
\label{fig:reach}
\end{figure}

The expected sensitivity of PADME is shown in Fig.~\ref{fig:reach}, for two different  number of positrons-on-target (POT). The more optimistic scenario is based on 2 years of data-taking with a 200 ns bunch length and a 30 Hz bunch rate. The more conservative scenario assumes 6 months of data-taking instead, or a 7.5 Hz effective bunch rate.

The next PADME data-taking run is expected to begin in early 2020. By then the beam-induced backgrounds should be mitigated, and the commissioning of sub-detectors complete. PADME will be ready to perform the main dark photon analysis as well as additional parallel searches. For example, PADME has sensitivity to axion-like particles (ALPs), through the $aee$ and $a\gamma\gamma$ couplings, and potentially to a subset of the parameter space available for inelastic dark matter. The concrete sensitivity to these extra models will be determined soon.

After the completion of the planned data-taking program at LNF, a few options for the future are currently being considered: PADME could move its detector to the Wilson Laboratory at Cornell, which has a 6 GeV positron beam energy, extending PADME's reach by a factor of 10 in dark photon mass. Moreover, a proposal to implement slow resonant extraction of positrons in the Cornell accelerator could stretch out the bunch lengths, reducing the in-time pileup on the detector and enhancing PADME's sensitivity to the kinetic mixing coupling. Another possibility is moving the detector to Jefferson Lab, where a 12 GeV positron beam would increase the mass reach by a factor of 20. All these potential avenues for future research are currently being assessed.

\section*{Acknowledgements}

This work is partly supported by the Italian Ministry of Foreign Affairs and International Cooperation (MAECI), CUP I86D16000060005.

\end{document}